\documentclass[prl,twocolumn,superscriptaddress,showpacs,amsmath,amssymb]{revtex4}
\usepackage{graphicx}
\usepackage{dcolumn}
\usepackage{bm}
\usepackage{color}

\newcommand{\beq}{\begin{eqnarray}}
\newcommand{\eeq}{\end{eqnarray}}

\begin{document}

\title{
Symmetry-Protected Majorana Fermions in Topological Crystalline
Superconductors:
\\
Theory and Application to Sr$_2$RuO$_4$
}

\author{Yuji Ueno}
\affiliation{Department of Applied Physics, Nagoya University,
Nagoya 464-8603, Japan}
\author{Ai Yamakage}
\affiliation{Department of Applied Physics, Nagoya University,
Nagoya 464-8603, Japan}
\author{Yukio Tanaka}
\affiliation{Department of Applied Physics, Nagoya University,
Nagoya 464-8603, Japan}
\author{Masatoshi Sato}
\email{msato@nuap.nagoya-u.ac.jp}
\affiliation{Department of Applied Physics, Nagoya University,
Nagoya 464-8603, Japan}
\date{\today}

\begin{abstract}
Crystal point group symmetry is shown to protect Majorana fermions
(MFs) in spinfull superconductors (SCs). 
We elucidate the condition necessary to obtain MFs
protected by the point group symmetry.
We argue that
superconductivity in Sr$_2$RuO$_4$ hosts a topological phase
transition to a topological crystalline SC,
 which accompanies a ${\bm d}$-vector rotation under a magnetic field
 along the $c$-axis.
Taking all three bands and spin-orbit interactions into account,
symmetry-protected MFs in the topological crystalline SC
are identified. Detection of such MFs provides evidence of the
${\bm d}$-vector rotation in Sr$_2$RuO$_4$ expected from Knight
shift measurements but not yet verified.
\end{abstract}

\pacs{
}


\maketitle

{\it Introduction --}
There has been a recent interest in the realization of Majorana
fermions (MFs) in topological superconductors (SCs) \cite{TSN12,
qi11, wilczek09}. While it is known that spin-triplet SCs can host
topological superconductivity \cite{RG00, Ivanov01, Sato09, Sato10,
FB10}, a recent breakthrough indicates that conventional $s$-wave
superconducting states may also support the topological phase in the
presence of spin-orbit interactions \cite{Sato03, fu08, STF09,
SLTS10, Alicea10}. The features of $s$-wave superconductivity and
its possible application to fault-tolerant topological quantum
computation have stimulated both theoretical and experimental
activities. The $s$-wave superconducting scheme has been applied to
a wide class of condensed matter systems 
\cite{STF10, Liu2012, Wei12, Zhai12, lutchyn10, ORO10, alicea2011}.

While these developments are based on topological classifications
using the general symmetries of time-reversal and charge conjugation
\cite{SRFL08}, systems often have other symmetries specific to their
structures such as translational, rotational, and point group
symmetries \cite{AM}. Interestingly, additional symmetries can give rise to a
nontrivial topology of the bulk wave functions and gapless states on
the boundaries \cite{Fu11, HPB11, TZV10, HLLDBF12,SMJZ13}. Although these
specific symmetries are microscopically sensitive to a small
disturbance, recent studies of topological crystalline insulators
have shown that if the symmetries are preserved on average, then the
existence of gapless boundary states is rather robust \cite{RKS12,
MBM12, FK12, FHEA12}. Therefore, it is expected that the symmetry-protected
topological phase can provide an alternative platform for realizing
MFs.

In this Letter, we clarify how the crystal point group symmetry may
protect the existence of MFs in SCs. 
We focus here on {\it spinfull SCs with
point group symmetry}. 
While the time-reversal symmetry may protect MFs in spinful SCs
\cite{QHRZ09, SRFL08, Roy08}, the
present theory also works even without the time-reversal invariance.
Our arguments are directly applicable to many
unconventional SCs, most of which are spinfull.

Using a representation of the gap function for the point group
symmetry, we first elucidate the condition necessary to obtain MFs
protected by the point group symmetry. We demonstrate that when this
condition is satisfied, spinfull SCs can be separated into a pair of
spinless topological SCs, each of which supports MFs. If the
conditions are not satisfied, however, the system reduces to a pair
of states that are topologically of the same class as quantum Hall
states, and thus they only support Dirac fermions at most.

As a concrete example, we apply these arguments to the
two-dimensional SC in Sr$_2$RuO$_4$. It is argued that Sr$_2$RuO$_4$
hosts a profound topological phase transition from a chiral topological SC to
a topological crystalline SC accompanied by ${\bm d}$-vector
rotation under a magnetic field parallel to the $c$-axis. We
identify the symmetry-protected MFs by taking all three bands and
spin-orbit interactions into account. Detection of the
symmetry-protected MFs provides a distinct signature of the
${\bm d}$-vector rotation in Sr$_2$RuO$_4$. Such a rotation is
expected from Knight shift measurements but is yet to be verified.

{\it Symmetry Protected Majorana Fermions --}
We first consider two-dimensional SCs, and later discuss the generalization to
three-dimensional SCs.
We begin with a description of two-dimensional SC based on the
Bogoliubov de Genne (BdG) Hamiltonian ${\cal H}=\sum_{{\bm
k}}\Psi^{\dagger}_{{\bm k}}{\cal H}({\bm k})\Psi_{\bm k}/2$:
\begin{eqnarray}
{\cal H}({\bm k})=\left(
\begin{array}{cc}
{\cal E}({\bm k}) & \Delta({\bm k})\\
\Delta^{\dagger}({\bm k}) & -{\cal E}^T(-{\bm k})
\end{array}
\right),
\Psi_{\bm k}=(c_{\bm k sl}, c^{\dagger}_{-{\bm k} sl})^{\rm t},
\label{eq:BdG}
\end{eqnarray}
where $c_{{\bm k}sl}$ is the annihilation operator of electrons with
momentum ${\bm k}=(k_x,k_y)$ and spin $s$, $l$ denotes the orbital
degrees of freedom of the electron, ${\cal E}({\bm k})$
is the Hamiltonian of the normal state, and $\Delta({\bm k})$ is the
gap function of the SC. Here, the spin ($s=\uparrow,\downarrow$) and
orbital ($l=1,\cdots, N$) indices are implicit in ${\cal E}({\bm
k})$ and $\Delta({\bm k})$. Both ${\cal E}({\bm k})$ and
$\Delta({\bm k})$ are $2N\times 2N$ matrices. Assuming that the
normal state has a mirror symmetry with respect to the $xy$-plane as $ {\cal
M}_{xy}{\cal E}({\bm k}){\cal M}_{xy}^{\dagger}={\cal E}({\bm k}), $
where ${\cal M}_{xy}$ is the unitary matrix of the mirror
reflection, we demonstrate how this mirror symmetry ensures
topologically stable MFs.

We note that the superconducting state retains the mirror symmetry
if the gap function $\Delta({\bm k})$ is even or odd under the
mirror reflection, $ {\cal M}_{xy}\Delta({\bm k}){\cal
M}_{xy}^{t}=\pm \Delta({\bm k}). $ In the former (latter) case,
${\cal H}({\bm k})$ is invariant under the mirror reflection
\begin{eqnarray}
\tilde{{\cal M}}_{xy}{\cal H}({\bm k})\tilde{{\cal M}}_{xy}^{\dagger}
={\cal H}({\bm k}),
\label{eq:mirrorBdG}
\end{eqnarray}
with $\tilde{\cal M}_{xy}=\tilde{{\cal M}}^{+}_{xy}$ ($\tilde{{\cal M}}_{xy}^{-}$) given by,
\begin{eqnarray}
\tilde{{\cal M}}_{xy}^{\pm}=\left(
\begin{array}{cc}
{\cal M}_{xy} & 0\\
0 & \pm {\cal M}_{xy}^{*}
\end{array}
\right).
\label{eq:mirror}
\end{eqnarray}
In both these cases, ${\cal H}({\bm k})$ commutes with either of
$\tilde{{\cal M}}^{\pm}_{xy}$ (which we refer to simply as
$\tilde{\cal M}_{xy}$ in the following). Therefore, ${\cal H}({\bm
k})$ is block diagonal in the diagonal basis of $\tilde{{\cal
M}}_{xy}$. Each block-diagonal subsector has a definite eigenvalue of
the matrix $\tilde{{\cal M}}_{xy}$. 

The mirror Chern number $\nu(\lambda)$ is defined as a Chern number
of the subsector; using the negative energy states
$|u^{\lambda}_n({\bm k})\rangle$ in the subsector with the
eigenvalue $\lambda$ of $\tilde{{\cal M}}_{xy}$, the gauge field in
the momentum space is introduced as ${\cal A}_{a}^{\lambda}({\bm
k})=i\sum_{E_n<0} \langle u^{\lambda}_n({\bm k})|\partial_{k_a}
u^{\lambda}_n({\bm k})\rangle$. The mirror Chern number is then
given by
\begin{eqnarray}
\nu(\lambda)=\frac{1}{2\pi}\int_{-\pi}^{\pi}\int_{-\pi}^{\pi} d k_x dk_y
 {\cal F}^{\lambda},
\end{eqnarray}
where ${\cal F}^{\lambda}$ is the field strength of ${\cal
A}_a^{\lambda}$, and the integration is performed over the first
Brillouin zone, $-\pi\le k_{x, y}\le \pi$.

The mirror Chern number introduced here is a natural generalization
of that used to characterize topological crystalline insulators
\cite{Fu11}. Therefore, as well as topological crystalline
insulators, the mirror Chern number ensures the existence of gapless
boundary states as
long as the mirror symmetry is preserved macroscopically. However,
as we show, there is an important difference: to stabilize the MFs
using the mirror symmetry, an additional requirement for
$\tilde{\cal M}_{xy}$ is needed.

This additional requirement originates from the symmetry specific to
SCs. Because of the self-conjugate property of the quasiparticle
field $\Psi_{\bm k}$ in Eq. (\ref{eq:BdG}),
\begin{eqnarray}
\Psi_{\bm k}={\cal C}\Psi_{-\bm k}^*,
\quad
{\cal C}=\left(
\begin{array}{cc}
0 &1 \\
1 &0
\end{array}
\right),
\end{eqnarray}
the BdG Hamiltonian has the special symmetry $ {\cal C}{\cal H}({\bm
k}){\cal C}^{\dagger}=-{\cal H}^*(-{\bm k});$ i.e., particle--hole
symmetry. The self-conjugate property of the particle--hole symmetry
is the origin of the Majorana nature of topologically protected
gapless states. Therefore, to obtain MFs in a subsector of the
system, particle--hole symmetry, which is closed in the subsector,
is essential. A subsector of ${\cal H}({\bm k})$ with a definite
eigenvalue of $\tilde{{\cal M}}_{xy}$, 
however, does not always have
particle--hole symmetry within the subsector because this symmetry
can exchange a pair of subsectors with different eigenvalues.
Thus, only subsectors with particle--hole symmetry can support
MFs that are protected by the mirror symmetry.

We now elucidate the condition for a subsector with an
eigenvalue $\lambda$ of $\tilde{\cal M}_{xy}$ 
to host its own
particle--hole symmetry.
Since the particle--hole symmetry maps a state $|u({\bm k})\rangle$
with the eigenvalue $\lambda$ to ${\cal C}|u^*(-{\bm k})\rangle$,
the condition is derived so that the mapped state ${\cal
C}|u^*(-{\bm k})\rangle$ has the same eigenvalue $\lambda$ as the
original one. We find that this leads to
\begin{eqnarray}
{\cal C}\tilde{\cal M}_{xy}{\cal C}^{\dagger}=\lambda^2\tilde{\cal
 M}_{xy}^*.
\label{eq:SPMFcond}
\end{eqnarray}
In other words, only for $\tilde{\cal M}_{xy}$ satisfying Eq.
(\ref{eq:SPMFcond}) one can obtain MFs protected by the mirror
symmetry.

\begin{figure}[t!]
\includegraphics[width=80mm]{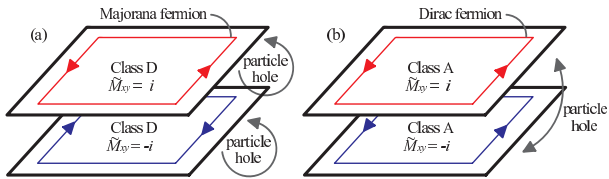}
\caption{Two possible realizations of particle--hole symmetry.}
\label{fig:1}
\end{figure}

Among the two possible mirror symmetries $\tilde{\cal M}_{xy}^{\pm}$
in Eq. (\ref{eq:mirror}), only $\tilde{\cal M}_{xy}^{-}$ satisfies
Eq. (\ref{eq:SPMFcond}). Indeed, from ${\cal
M}_{xy}^2=-1=\lambda^2$, one can show that $\lambda=\pm i$ and
${\cal C}\tilde{\cal M}_{xy}^{\pm}{\cal C}^{\dagger} =\mp
\lambda^2 \tilde{\cal M}_{xy}^{\pm*}$. 
These are consistent with Eq.
(\ref{eq:SPMFcond}) only for $\tilde{\cal M}_{xy}^{-}$, which means
that when the gap function is odd under the mirror reflection
${\cal M}_{xy}\Delta({\bm k}){\cal M}_{xy}^{t}=-\Delta({\bm k})$,
the subsector with the eigenvalue $\lambda$ supports its own
particle--hole symmetry [Fig. \ref{fig:1}(a)]. However, when the gap
function is even under the mirror reflection ${\cal
M}_{xy}\Delta({\bm k}){\cal M}_{xy}^{t}=\Delta({\bm k})$, the
particle--hole symmetry of the original Hamiltonian merely
interchanges two different subsectors, and thus each subsector does
not have particle--hole symmetry [Fig. \ref{fig:1}(b)].

In the latter case,
we find that the subsectors belong to class A of the topological
classification \cite{SRFL08} since there is no particle--hole
symmetry. That is, they are topologically the same as quantum Hall
states, which also belong to class A \cite{footnote1}.
Therefore, even when there are topologically protected states
ensured by the mirror Chern number, these states reduce to Dirac
fermions as is the case for quantum Hall states.

In contrast, MFs can be realized in the former case: each subsector
belongs to class D of the topological classification \cite{SRFL08}
because of its own particle--hole symmetry. Consequently, as a class
D topological phase, the mirror subsector can host MFs. The
topological invariant for class D is 0 for three dimensions, ${\bm
Z}$ for two dimensions, and ${\bm Z}_2$ for one dimension.
Therefore, in addition to the mirror Chern number $\nu(\lambda)$, we
can define the one-dimensional (1D) ${\bm Z}_2$ invariants
$\nu_x(\lambda)$ and $\nu_y(\lambda)$ as \cite{Sato10}
\begin{eqnarray}
\nu_x(\lambda)=\frac{1}{\pi}\int_{-\pi}^{\pi}dk_y {\cal A}_y^{\lambda}(\pi,k_y)
\quad \mbox{mod $2$},
\end{eqnarray}
which we refer to as mirror ${\bm Z}_2$ invariants ($\nu_y(\lambda)$
is defined in a similar manner). The 1D mirror ${\bm Z}_2$
invariants are well defined even for a system without time-reversal
invariance, and thus they are very different from the ${\bm Z}_2$
invariants for topological (crystalline) insulators.

The topological invariants $\nu({\lambda})$, $\nu_x({\lambda})$, and
$\nu_y({\lambda})$ characterize the MFs in the mirror subsector; the
mirror Chern number ensures the existence of chiral edge MFs in the
mirror subsector. Using a similar argument to that given in Refs.
\cite{Roy10} and \cite{TK10}, we can also show that if the mirror
Chern number is odd, then the Majorana zero-energy states exist on a
vortex. Furthermore, the 1D mirror ${\bm Z}_2$ invariants guarantee
the existence of Majorana zero-energy bound states localized on a
dislocation, if the Burgers vector ${\bm B}$ characterizing the
dislocation satisfies $ {\bm B}\cdot {\bm G}^{\lambda}=1 $ mod $2$
\cite{AN12}. Here, ${\bm
G}^{\lambda}=\frac{1}{2\pi}(\nu_x(\lambda){\bm
b}_x+\nu_y(\lambda){\bm b}_y)$ with ${\bm b}_x$ and ${\bm b}_y$
being the reciprocal lattice vectors in the $x$- and $y$-directions.
All these MFs are topologically stable so long as the mirror
symmetry is preserved macroscopically beyond the scale of the
coherence length.

A generalization of these results to three-dimensional SCs is
fairly straightforward. In three dimensions, Eq.
(\ref{eq:mirrorBdG}) is replaced with $ \tilde{\cal M}_{xy}{\cal
H}(k_x,k_y,k_z)\tilde{\cal M}_{xy}^{\dagger} ={\cal H}(k_x,k_y,
-k_z).$ If we then consider the mirror invariant $k_xk_y$-planes
with $k_z=0$ or $\pi$, the same argument as per the two-dimensional
case applies, and again symmetry-protected MFs can be realized when
the gap function is odd under the mirror reflection. We can also
generalize our arguments to other point group symmetries. 
For
instance, 
like topological
crystalline insulators, we can introduce topological invariants
using discrete rotation symmetries, but to stabilize the MFs,
additional requirements similar to those in Eq. (\ref{eq:SPMFcond})
are needed.

{\it Application to Sr$_2$RuO$_4$--}
To illustrate the general arguments above, we apply our results to
the two-dimensional spin-triplet SC Sr$_2$RuO$_4$ \cite{MM03,
MKNYI12}. Sr$_2$RuO$_4$ has a tetragonal structure with the crystal
point group symmetry $D_{4h}$ and, in particular, is invariant under
a mirror reflection with respect to the $xy$-plane. We consider the
superconducting state of Sr$_2$RuO$_4$ under magnetic fields in the
$z$-direction. In this case, the crystal point group symmetry
reduces to $C_{4h}$, but the system is still invariant under the
mirror reflection ${\cal M}_{xy}$. According to the representation
theory of $C_{4h}$, the possible spin-triplet gap functions are
classified as in Table \ref{table:gap function}.

For zero or weak magnetic fields, a number of experiments support
time-reversal breaking chiral $p_x+ip_y$ superconductivity in
Sr$_2$RuO$_4$, which belongs to the $E_{u}$ representation in Table
\ref{table:gap function} \cite{MM03, MKNYI12}. The ${\bm d}$-vector
is aligned along the $z$-direction, and the gap function,
$\Delta({\bm k})=i{\bm d}({\bm k})\cdot{\bm \sigma}\sigma_y$, is even
under the mirror reflection. For magnetic fields greater than 20 mT,
Knight shift measurements suggest that the ${\bm d}$-vector is
parallel to the $xy$-plane, consistent with the $A_u$ or $B_u$
representation in Table \ref{table:gap function} \cite{MIKMM04}.
Thus, the gap function is helical and odd under the mirror
reflection. For consistency with both experimental results, there
should be a phase transition at a critical magnetic field $H_{z \rm
c}$ at which the ${\bm d}$-vector rotates from the $z$-direction to
the $xy$-plane, the chiral superconducting state becomes helical,
and the mirror parity of the gap function changes from even to odd.

\begin{table}
\begin{ruledtabular}
\caption{Possible ${\bm d}$-vector states of Sr$_2$RuO$_4$ under
magnetic fields in the $z$-direction. The representation $\Gamma$,
${\bm d}$-vector, parity of the gap function $\Delta=i{\bm d}\cdot{\bm
 \sigma}\sigma_y$ under the mirror
reflection ${\cal M}_{xy}$, topological class of the subsector with
a definite eigenvalue of $\tilde{\cal M}_{xy}$, and mirror
topological numbers $\nu$, $\nu_x$, and $\nu_y$ are summarized. Note
that the mirror ${\bm Z}_2$ numbers are not defined in the $E_u$
representation.} \begin{tabular}{ccccc} $\Gamma$ & ${\bm
d}$-vector & ${\cal M}_{xy}$ & subsector & $[\nu(\lambda),
\nu_x(\lambda),
  \nu_y(\lambda)]$
\\ \hline
$A_{u}$ & $\hat{\bm x}\sin k_x+\hat{\bm y}\sin k_y$ & odd  & class
D & $[\pm 1,1,1]_{\lambda=\pm i}$
\\
  & $\hat{\bm x}\sin k_y-\hat{\bm y}\sin k_x$
&   &
\\ $B_{u}$ & $\hat{\bm x}\sin k_x-\hat{\bm y}\sin k_y$
& odd  & class D & $[\mp 1,1,1]_{\lambda=\pm i}$
\\
 & $\hat{\bm x}\sin k_y+\hat{\bm y}\sin k_x$
&  &
\\
$E_{u}$  & $\hat{\bm z}(\sin k_x+ i \sin k_y)$
& even  & class A & $[1,-, -]_{\lambda=\pm i}$
\\
 \end{tabular}
\label{table:gap function}
\end{ruledtabular}
 \end{table}

It is known that the chiral $p_x+ip_y$ superconducting state in the
low-field phase hosts topological superconductivity \cite{Volovik97,
GI98}. Taking the spin degrees of freedom into account, the Chern
number of this state is evaluated as $\nu_{\rm Ch}=2$, and thus a
pair of topologically protected chiral fermions exist on the
boundary \cite{YTK97, HS98, MS99, FMS01, SHY02, KKKFYTM11, IWS12}. However,
our argument implies that it is impossible to detect their Majorana
character if ${\bm d}\parallel \hat{\bm z}$. 
Since the gap function is even under the mirror
reflection, this state can be separated into a pair of subsectors
belonging to class A. In each subsector, the paired MFs, $\gamma_1$
and $\gamma_2$, are recast into a single Dirac fermion
$\psi=\gamma_1+i\gamma_2$, and thus their dynamics are reduced to
those of the Dirac fermion.

In contrast, the helical superconducting state in the high-field
phase supports MFs. The ordinary Chern number $\nu_{\rm Ch}$ is
found to be zero in this phase, and so this phase is not topological
in terms of the topological periodic table \cite{SRFL08}.
Nevertheless, since the gap function is odd under the mirror
reflection, the mirror Chern number and the 1D mirror ${\bm Z}_2$
invariants can be introduced. Using a standard adiabatic deformation
method for the Hamiltonian, the mirror topological invariants can be
calculated, and we find that they are nonzero. The obtained results
are summarized in Table \ref{table:gap function}. 
(For details, see Sec.S2 in Supplementary Material.)
From the
bulk--boundary correspondence, one can conclude that MFs protected
by the mirror symmetry exist. Since the mirror symmetry is essential
for hosting the topological phase, the high-field phase realizes a
topological crystalline SC\cite{TH12}.

To illustrate these results, we examine the quasiparticle states of
Sr$_2$RuO$_4$ using the following model Hamiltonian on the square
lattice \cite{NS00}. The conduction bands of Sr$_2$RuO$_4$ consist
of three 4d-$t_{2g}$ Ru orbitals, $d_{xz}$, $d_{yz}$, and $d_{xy}$,
which we label as $l=1,2,$ and 3, respectively.
The superconducting state of Sr$_2$RuO$_4$ is described by the BdG
Hamiltonian ${\cal H}={\cal H}_{\rm kin}+{\cal H}_{\rm so}+{\cal
H}_{\rm pair}$, in which
\begin{eqnarray}
&&{\cal H}_{\rm kin}=\sum_{{\bm k}s}
(c^{\dagger}_{{\bm k}s1}, c^{\dagger}_{{\bm k}s2}, c^{\dagger}_{{\bm k}s3})
\left(
\begin{array}{ccc}
\varepsilon_{{\bm k}1} & g_{\bm k}& 0\\
g_{\bm k} &\varepsilon_{{\bm k}2}& 0 \\
0 & 0 & \varepsilon_{{\bm k}3}
\end{array}
\right)
\left(
\begin{array}{c}
c_{{\bm k}s1} \\
c_{{\bm k}s2} \\
c_{{\bm k}s3}
\end{array}
\right),
\nonumber\\
&&{\cal H}_{\rm so}=i\lambda
\sum_{lmn}\epsilon_{lmn}
\sum_{{\bm
 k}ss'}c_{{\bm k}sl}^{\dagger}c_{{\bm k}s'm}\sigma_{ss'}^n,
\nonumber\\
&&{\cal H}_{\rm pair}=\frac{1}{2}\sum_{{\bm k}lss'}
\hat{\Delta}^{l}_{ss'}({\bm k})c^{\dagger}_{{\bm  k}sl}c^{\dagger}_{-{\bm k}s'l}
+{\rm h.c.}
\label{eq:BdGSr2RuO4}
\end{eqnarray}
where
$\varepsilon_{{\bm k}1}=-2t_1\cos k_y-\mu$, $\varepsilon_{{\bm
k}2}=-2t_1 \cos k_x-\mu$, and $\varepsilon_{{\bm k}3}=-2t_2(\cos
k_x+\cos k_y)-4t_3\cos k_x\cos k_y-\mu'$ are the kinetic terms of
orbital 1, 2, and 3, respectively, $g_{\bm k}=-4t_4\sin k_x\sin k_y$
is the hybridization of orbital 1 and 2, and $\hat{\Delta}^{l}({\bm
k})=i\Delta^{l}{\bm d}({\bm k})\cdot {\bm \sigma} {\sigma}_y$ is the
gap function with ${\bm d}({\bm k})$ as defined in Table
\ref{table:gap function}.
Here, $\sigma^n$ is the Pauli matrix, and $\epsilon_{lmn}$ is the
completely antisymmetric tensor. In the $(c_{{\bm k}s1},c_{{\bm
k}s2}, c_{{\bm k}s3})$ basis, ${\cal M}_{xy}$ is given by ${\cal
M}_{xy}={\rm diag}(-i\sigma_z, -i\sigma_z, i\sigma_z)$. For
definiteness, we assume ${\bm d}({\bm k})=\hat{\bm x}\sin
k_y-\hat{\bm y}\sin k_x$ in the high-field phase, but a
qualitatively similar result is obtained as long as the ${\bm
d}$-vector is parallel to the $xy$-plane.

\begin{figure}[t!]
\includegraphics[width=80mm]{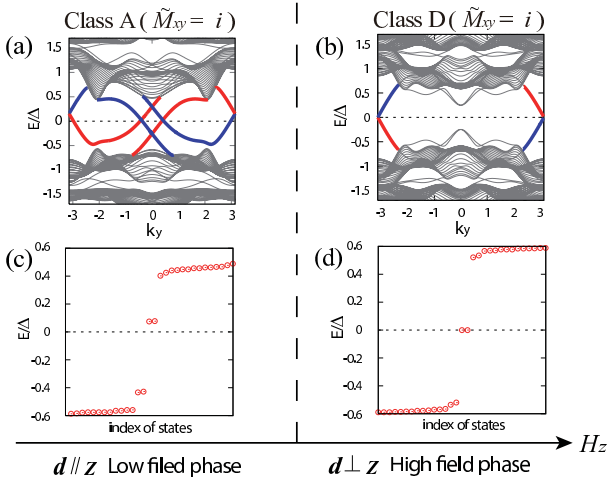}
\caption{Topological gapless states in the low- and high-field
phases of Sr$_2$RuO$_4$. (a) and (b) Edge states and (c) and (d)
bound states of the edge dislocation. In (a) and (b), the red
(blue) lines represent edge states localized at $x=0$ ($x=L$).
Results are shown for the $\tilde{\cal M}_{xy}=i$ sector. The ${\bm
d}$-vector is chosen as ${\bm d}({\bm k})=\hat{\bm
 z}(\sin k_x+i\sin k_y)$ in the low-field phase and as ${\bm d}({\bm
 k})=\hat{\bm x}\sin k_y-\hat{\bm y}\sin k_x$ in the high-field phase.
We take $t_1=t_2=0.5$, $t_3=0.2$, $t_4=0.1$, $\mu=-0.2$,
$\mu'=-0.2$, $\mu_{\rm B}H_z=0.1$, $\lambda=0.3$, and
$\Delta^{l}=0.6$.} \label{fig:2}
\end{figure}

To examine the edge states of Sr$_2$RuO$_4$, we numerically
diagonalize the BdG Hamiltonian Eq. (\ref{eq:BdGSr2RuO4}) in the
coordinate space with open boundaries at $x=0$ and $x=L=30$. Using
the mirror symmetry, we separate the quasiparticle spectra into two
subsectors with different eigenvalues of $\tilde{\cal M}_{xy}$.
Figures \ref{fig:2}(a) and (b) illustrate the quasiparticle spectra
in the $\tilde{\cal M}_{xy}=i$ sector. A similar result is obtained
in the $\tilde{\cal M}_{xy}=-i$ sector. Gapless edge states for both
the low-field phase [Fig. \ref{fig:2} (a)] and the high-field phase
[Fig. \ref{fig:2} (b)] can be seen clearly, which is consistent with
the nonzero mirror Chern numbers in both phases. 
The degeneracy of these edge states
at $k_y=0$ \cite{YTK97, HS98, MS99, FMS01,
SHY02} is resolved by the multi-orbital spin-orbit interaction.
Note, however, that
there is an essential difference between the two phases. While the
quasiparticle spectrum in the high-field phase is particle--hole
symmetric as $E(k_y)=-E(-k_y)$, that in the low-field phase is not.
This indicates clearly that only the high-field phase supports edge
MFs. Note also that the particle--hole symmetry in the low-field
phase is recovered if it is considered in conjunction with the $\tilde{\cal
M}_{xy}=-i$ sector.

Here we would like to mention that the existence of gapless states in
Fig.\ref{fig:2} (b) is consistent with the Ising behavior of helical
MFs \cite{SF09, CZ09, Volovik10, SFN10}: Helical
MFs remain gapless except under a magnetic field along a special
direction determined by the ${\bm d}$-vector.
As well as Ref.\cite{MSM12}, the present result reveals that
symmetry protection is essential for the Majorana Ising character.

The bound states of the edge dislocations are shown in Figs.
\ref{fig:2}(c) and (d). 
The
energy spectra were calculated on $30\times 30$ unit cell system
with periodic boundary conditions in the $x$- and $y$-directions.
Two edge dislocations with the Burgers
vector ${\bm B}=\pm \hat{\bm e}_x$, separated by half the length of
the system size, are considered. 
(See Sec.S3 in Supplementary Material.)
In the high-field phase, we have
${\bm G}^{\lambda}=\frac{1}{2\pi}({\bm b}_x+{\bm b}_y)$ from Table
\ref{table:gap function}, so ${\bm G}^{\lambda}\cdot {\bm B}=1$,
which predicts the existence of a zero mode in the dislocation. 
The high-field phase can be seen to support zero energy states,
which are localized on the edge dislocations and are two-fold
degenerate because there are two dislocations in the system.
Therefore, there is a single Majorana zero mode localized on each
edge dislocation.

In the same way as ordinary MFs, these symmetry-protected MFs can be
identified by tunneling spectroscopy. In particular, the detection
of the zero mode in the dislocation gives a strong signature of the
${\bm d}$-vector rotation in the high-field phase; if the ${\bm
d}$-vector rotation does not occur, the bound state in the
dislocation must have a gap of $O(\mu_{\rm B}H_z)$. Thus, the
observation of the zero energy state is a distinct signal of the
${\bm d}$-vector rotation. The change in the edge-mode spectra can
also provide details of the tunneling conductance. In addition, in
each sector of a definite 
eigenvalue of $\tilde{\cal M}_{xy}$, the
symmetry-protected MFs behave like ordinary MFs and retain the
unique features specific to MFs \cite{BD07, JAB08, akhmerov09,
law09, TYN09, LTYSN10, alicea12, Beenakker12}.

{\it Discussions --}
The symmetry-protected MFs discussed here are applicable to many
unconventional SC; the same arguments apply equally to
Cu$_x$Bi$_2$Se$_3$ \cite{Hor10, Wray10, KSKYTSA11, FB10, YYST12,
HF12} and UPt$_3$ \cite{RT02, Machida12} as both these crystals have
a mirror plane. Details about these systems will be reported
elsewhere.

The authors are grateful to S. Kashiwaya
for fruitful discussions. This work was supported by the JSPS
(Nos.~2074023303, 2134010303, and 22540383) and KAKENHI
Grants-in-Aid (Nos.~22103002 and 22103005) from MEXT.

\bibliography{Sr2RuO4}

\begin{thebibliography}{71}
\bibitem[{foo()}]{footnote1}
\bibinfo{note}{This does not imply the time-reversal breaking of the gap
  function while the time-reversal symmetry is broken in the quantum Hall
  state. Because the time-reversal operation flips the spin, the mirror
  subsector does not have the time-reversal symmetry even if the whole system
  is time-reversal invariant.}
\end{thebibliography}

\clearpage
\onecolumngrid

\renewcommand{\thefigure}{S\arabic{figure}} 

\renewcommand{\thesection}{S\arabic{section}.}

\renewcommand{\theequation}{S.\arabic{equation}}

\setcounter{figure}{0}
\setcounter{equation}{0}


\begin{flushleft} 
{\Large {\bf Supplementary Material}}
\end{flushleft}

\baselineskip24pt

\begin{flushleft} 
{\bf S1. Parity of the gap function under the mirror reflection 
${\cal M}_{xy}$ }
\end{flushleft} 

Under the mirror reflection with respect to the $xy$-plane, the gap
function transforms as 
\begin{eqnarray}
\hat{\Delta}({\bm k})\rightarrow {\cal M}_{xy}\hat{\Delta}(k_x, k_y,
-k_z){\cal M}^t_{xy},
\label{eq:3dmirror}
\end{eqnarray}
with ${\bm k}=(k_x,k_y,k_z)$ and ${\cal M}_{xy}=\pm i\sigma_z$.
If the gap function satisfies
$
{\cal M}_{xy}\hat{\Delta}(k_x, k_y, -k_z){\cal
 M}^t_{xy}=\hat{\Delta}({\bm k})
$ ($
{\cal M}_{xy}\hat{\Delta}(k_x, k_y, -k_z){\cal
 M}^t_{xy}=-\hat{\Delta}({\bm k}) 
$),
the parity of the gap function under the mirror reflection is even (odd).
In general, the mirror parity is determined completely by the
representation of the gap function under the point group symmetry.

For example, consider a gap function that belongs to the
trivial representation of an $s$-wave gap function $\hat{\Delta}_{\rm
s} ({\bm k})=i\psi\sigma_y$.  
In this case, 
Eq.(\ref{eq:3dmirror}) leads ${\cal M}_{xy}\hat{\Delta}_{\rm s}(k_x, k_y,
-k_z){\cal M}^t_{xy}=\hat{\Delta}_{\rm s}({\bm k})$, 
which implies that the mirror parity of the gap function in the trivial
representation  is always even.
 
For a two dimensional spin-triplet gap function $\hat{\Delta}({\bm
k})=i{\bm d}({\bm k}){\bm \sigma}\sigma_y$ with ${\bm k}=(k_x,k_y)$, we
can relate the mirror parity to the orientation of the ${\bm d}$-vector.   
In this case, the right hand side of Eq.(\ref{eq:3dmirror}) becomes
\begin{eqnarray}
-id_x(k_x,k_y)\sigma_x\sigma_y-id_y(k_x,k_y)\sigma_y\sigma_y 
+id_z(k_x,k_y)\sigma_z\sigma_y, 
\end{eqnarray}
and thus if the ${\bm d}$-vector is normal (parallel) to the
$z$-direction, the mirror parity is odd (even), and vice versa.
Therefore, for a two dimensional spin-triplet superconductor like
Sr$_2$RuO$_4$, the mirror parity is determined by the orientation of the
${\bm d}$-vector and does not depend on other details of the gap function.

\begin{flushleft} 
{\bf S2. Mirror topological numbers of Sr$_2$RuO$_4$}
\end{flushleft} 

In this section, we calculate the mirror topological numbers of Sr$_2$RuO$_4$.
To calculate the mirror topological numbers, we adiabatically turn off
the spin-orbit interaction $\lambda$ and the inter-orbit hopping $t_4$
of the Hamiltonian
Eq.(8) without gap closing, by changing other parameters of the
Hamiltonian slightly if necessary.
This process does not change the mirror topological numbers since they
can change only when the gap closes.
In the absence of the spin-orbit interaction and the inter-orbit
coupling, the three orbitals of Sr$_2$RuO$_4$ are decoupled to each other, 
and thus the calculation of the mirror topological numbers is simplified.

First, we calculate the mirror Chern number.
Among the three orbitals of Sr$_2$RuO$_4$, only the $d_{xy}$ orbital
contributes to the mirror Chern number 
since other two orbitals reduce
to be one-dimensional in the absence of the
inter-orbit coupling.
The BdG Hamiltonian of the $d_{xy}$ orbital is given by
\begin{eqnarray}
{\cal H}_{3}=\frac{1}{2}\sum_{{\bm k}ss'}
(c_{{\bm k}s 3}^{\dagger}, c_{-{\bm k}s 3})
\left(
\begin{array}{cc}
\varepsilon_{{\bm k}3}-\mu_{\rm  B}H_z\sigma_{z} 
& \hat{\Delta}^3({\bm k})
\\
\hat{\Delta}^{3\dagger}({\bm k})
& -\varepsilon_{-{\bm k}3}+\mu_{\rm B}H_z\sigma_{z} 
\end{array}
\right)_{ss'} 
\left(
\begin{array}{c}
c_{{\bm k}s'3} \\
c^{\dagger}_{-{\bm k}s' 3}
\end{array}
\right).
\end{eqnarray}
To calculate the mirror Chern number, we block diagonalize
the above Hamiltonian of the $d_{xy}$ orbital in the diagonal basis of
$\tilde{\cal M}_{xy}$. 
Each sector of the block diagonal Hamiltonian is a $2\times 2$ matrix
with the following form,
\begin{eqnarray}
{\cal H}_{\lambda}({\bm k})=\sum_{\mu=0}^{3}h_{\mu;\lambda}({\bm k})\sigma_{\mu},
\end{eqnarray}
where $\sigma_{\mu}=({\bm 1}, \sigma_a)$ $(a=1,2,3)$ is the 4-component Pauli
matrix, and $\lambda$ denotes the eigenvalue of $\tilde{\cal M}_{xy}$.
For the $2\times 2$ Hamiltonian, 
the mirror Chern number $\nu(\lambda)$ is recast into 
\begin{eqnarray}
\nu(\lambda)=\frac{1}{8\pi}\int_{-\pi}^{\pi}\int_{-\pi}^{\pi}dk_xdk_y
\epsilon^{ij}\epsilon^{abc}
\hat{h}_{a;\lambda}({\bm k})\partial_{k_i}
\hat{h}_{b;\lambda}({\bm k})\partial_{k_j}\hat{h}_{c;\lambda}({\bm k}),
 \end{eqnarray}
with $\hat{h}_{a;\lambda}=h_{a;\lambda}/\sqrt{\sum_{b=1}^3
h^2_{b;\lambda}}$.
Here the summations for $i,j=x,y$ and $a,b,c=1,2,3$ are implicit. 
The last equation is evaluated as [1]
\begin{eqnarray}
\nu(\lambda)=\frac{1}{2}\sum_{h_{1;\lambda}({\bm k})=h_{2;\lambda}({\bm k})=0}
{\rm sgn}[h_{3;\lambda}({\bm k})]
{\rm sgn}[\det\partial_{k_i}h_{j;\lambda}({\bm k})],
\end{eqnarray}
where the summation is taken for ${\bm k}$ satisfying
$h_{1;\lambda}({\bm k})=h_{2;\lambda}({\bm k})=0$, and
\begin{eqnarray}
{\rm det}\partial_{k_i}h_{j;\lambda}({\bm k})=
{\rm det}\left(
\begin{array}{cc}
\partial_{k_x}h_{1;\lambda}({\bm k})  &\partial_{k_x}h_{2;\lambda}({\bm k})  \\
\partial_{k_y}h_{1;\lambda}({\bm k}) &\partial_{k_y}h_{2;\lambda}({\bm k}) 
\end{array}
\right).
\end{eqnarray}
This formula leads to the results summarized in Table I.

Now calculate the mirror ${\bm Z}_2$ invariants for mirror even gap functions.
By using the technique developed in Ref.[7], it is found that the
mirror ${\bm Z}_2$ invariant is directly related to the Fermi surface
as
\begin{eqnarray}
(-1)^{\nu_x(\lambda)}=\prod_{l}{\rm sgn}\varepsilon_{{\bm k}=(\pi,0)l}
\,{\rm sgn}\varepsilon_{{\bm k}=(\pi,\pi)l}, 
\quad
(-1)^{\nu_y(\lambda)}=\prod_{l}{\rm sgn}\varepsilon_{{\bm k}=(0,\pi)l}
\,{\rm sgn}\varepsilon_{{\bm k}=(\pi,\pi)l},
\end{eqnarray}
which yields the results in Table I.

\begin{flushleft} 
{\bf S3. Model of edge dislocations}
\end{flushleft} 

\begin{figure}[h]
\begin{center}
\includegraphics[width=6cm]{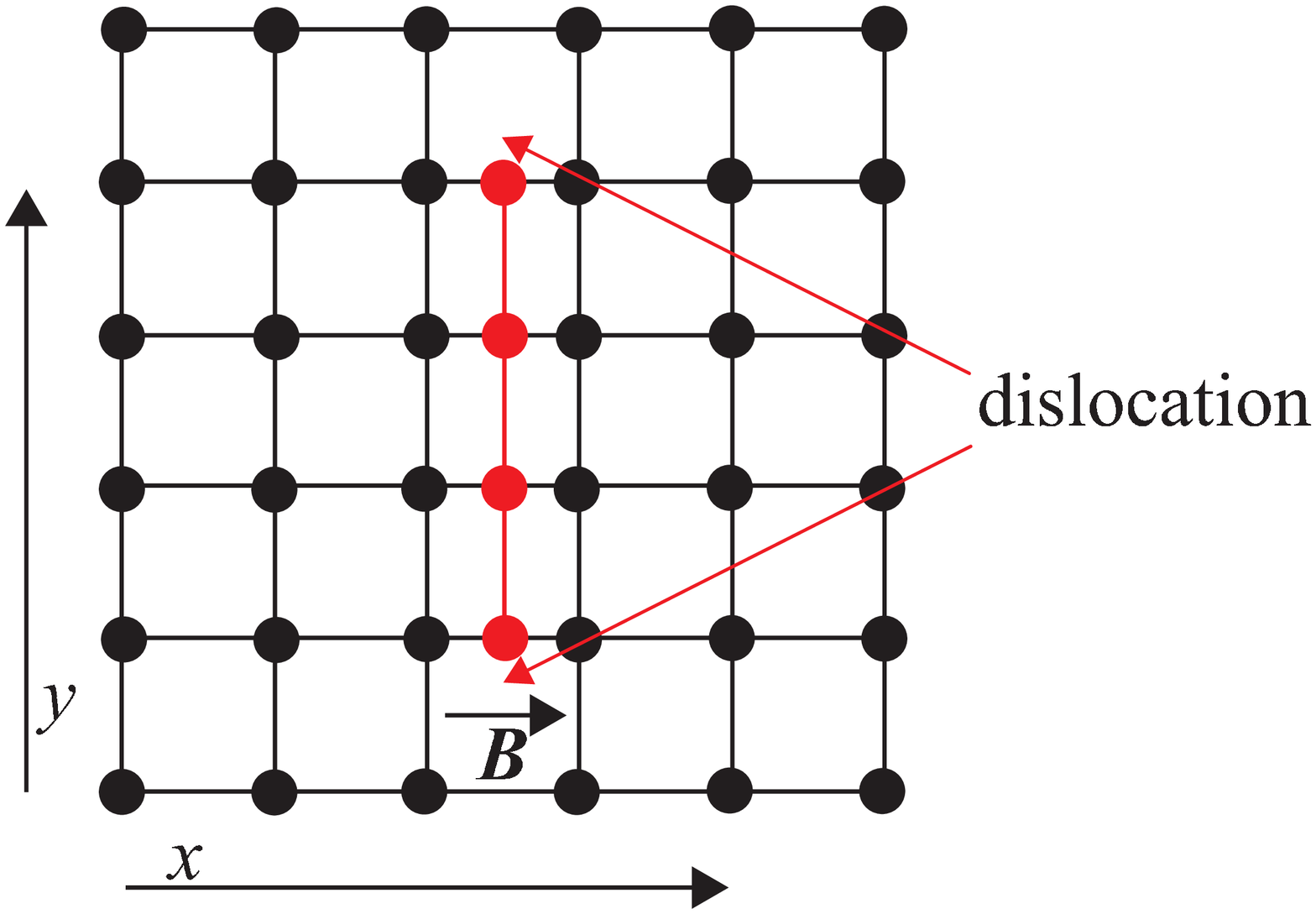}
\caption{Model of edge dislocations. Additional links are inserted to
 create edge dislocation with the Burgers vector ${\bm B}=\pm \hat{\bm e}_x$. } 
\label{fig:S1}
\end{center}
\end{figure}

\begin{figure}[h]
\begin{center}
\includegraphics[width=11cm]{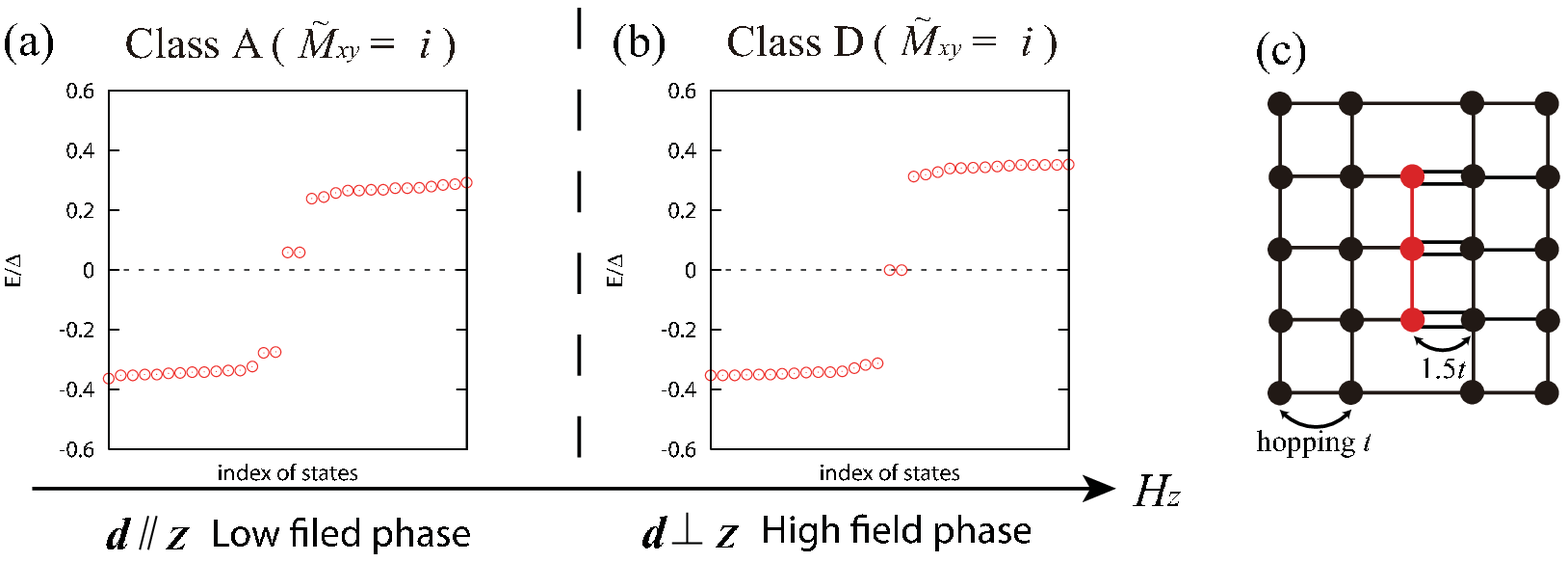}
\caption{(a) and (b) Bound states of the edge dislocations with deformed
 model parameters near the dislocations.
The ${\bm d}$-vector is chosen as ${\bm d}({\bm k})=\hat{\bm z}(\sin
 k_x+i\sin k_y)$ in the low-field phase and as ${\bm d}({\bm
 k})=\hat{\bm x}\sin k_y-\hat{\bm y}\sin k_x$ in the high-field phase. 
(c) Deformed edge dislocations. 
On the double lines in Fig.\ref{fig:S2}(c),
we take $t_1=t_2=0.75$, $t_3=0.3$, $t_4=0.15$ and $\Delta^l=0.9$.
In other regions, the model parameters are the same as those used in
 Fig. 2. 
} 
\label{fig:S2}
\end{center}
\end{figure}

We explain our model for edge dislocations used in the
 calculation of Figs. 2 (c) and (d).
As illustrated in Fig. \ref{fig:S1}, we place two edge dislocations with the
 Burgers vector ${\bm B}=\pm \hat{\bm e}_{x}$ and diagonalize the
 Hamiltonian Eq.(8) in the coordinate space. 
In actual calculation, we have used $30\times30$ unit cell system with
the periodic boundary condition and
 the two dislocations are separated by 15 unit length. 
Figures 2 (c) and (d) show the resultant eigen energies. 
While we have assumed that the parameters of the Hamiltonian near the
 dislocations are the same as those in the bulk, for
 simplicity, our result does not depend on the details:
In Fig.\ref{fig:S2}, we show the quasiparticle spectra of a deformed edge
 dislocation where the double links near the dislocation in Fig.\ref{fig:S2}
 (c) take different model parameters than those in the bulk.
Figures \ref{fig:S2} (a) and (b) clearly indicate that the qualitative
 behaviors are the same as those in Figs. 2(c) and (d), respectively.

\begin{flushleft} 
{\bf S4. Robustness of mirror topological phase in Sr$_2$RuO$_4$}
\end{flushleft}

In this letter, we have assumed conventional chiral $p$-wave gap
function in the low field phase of
Sr$_2$RuO$_4$. 
However, our qualitative results rarely depend on the
details of the gap function as far as the gap function is spin-triplet
and the orientation of the ${\bm d}$-vector is the same:
Since the parity of the gap function under the mirror reflection is the same
as far as the orientation of the ${\bm d}$-vector is the same, as we
pointed out in Sec.S1, the topological class of the mirror subsectors is
also the same.
Furthermore, for spin-triplet gap functions, one can show that the
mirror ${\bm Z}_2$ invariant (for ${\bm d}\perp \hat{\bm z}$) and the
parity of the mirror Chern number 
are the same as far as the 
Fermi surface topology in the normal state is the same [7].
This means that the mirror topological numbers are non-trivial,
irrespectively of details of the gap function for Sr$_2$RuO$_4$. 

\begin{figure}[h]
\begin{center}
\includegraphics[width=5cm]{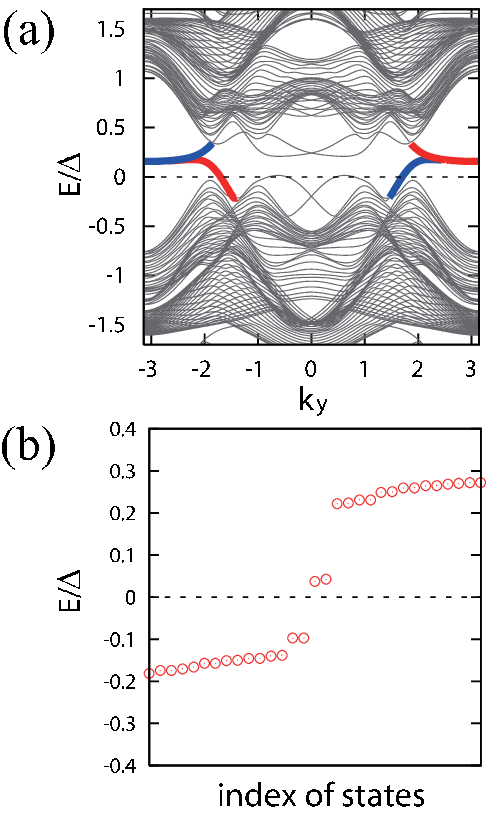}
\caption{Topological gapless states for the gap function
 Eqs.(\ref{eq:rkk1}) and (\ref{eq:rkk2}).
(a) Edge state and (b) bound state of the edge dislocation.
In (a), the red (blue) lines represent edge states localized at $x=0$ $(x=L)$.
The model parameters are the same as those of Fig.2. } 
\label{fig:S3}
\end{center}
\end{figure}

For comparison, Figs. \ref{fig:S3} (a) and (b) show the edge state and the bound
state of the edge dislocation, respectively.
The gap function is taken to be
a recent proposed one in the low field phase \cite{RKK10},
\begin{eqnarray}
\hat{\Delta}^l({\bm k})=i\Delta^{l} {\bm d}^{l}({\bm k})\cdot{\bm \sigma}\sigma_y,
\label{eq:rkk1}
\end{eqnarray}
with 
\begin{eqnarray}
{\bm d}^1({\bm k})=i\sin k_y\cos k_x \hat{\bm z},
\quad 
{\bm d}^2({\bm k})=\sin k_x\cos k_y \hat{\bm z}, 
\quad
{\bm d}^3({\bm k})=(\sin k_x+i\sin k_y) \hat{\bm z}. 
\label{eq:rkk2}
\end{eqnarray}
These spectra are indeed qualitatively the same as those in Figs. 2
(a) and (c).

\begin{figure}[h]
\begin{center}
\includegraphics[width=5cm]{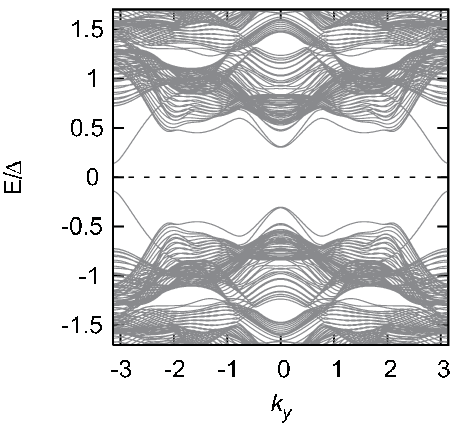}
\caption{ Edge state in Fig.2 (b) under a magnetic field along the
 $y$-direction which breaks the mirror symmetry with respect
 to the $xy$-plane. 
Here we show the quasiparticle spectrum of the total system, not that in
 a mirror subsector, since the mirror symmetry is broken.  
We take $\mu_{\rm B}H_y=0.1$ and $\mu_{\rm B}H_z=0$. 
Other model parameters are the same as those of Fig.2 (b). } 
\label{fig:S4}
\end{center}
\end{figure}

Our consideration also suggests that the gapless state in Fig. 2(b) can be
gapful if one breaks the mirror symmetry with respect to the $xy$-plane.
This is indeed the case as is illustrated in Fig.\ref{fig:S4}.

\begin{flushleft} 
{\bf S5. Effects of the spin-orbit interaction on edge
 modes in Sr$_2$RuO$_4$. }
\end{flushleft} 
While edge modes of Sr$_2$RuO$_4$ have been studied intensively[43-49],
the spin-orbit interaction had been ignored except for a recent
work by Imai et al. [49], where the spin-orbit interaction was taken into
account partially for the $\alpha$ and $\beta$-bands.
In the present letter, we consider the spin-orbit interaction for full
three $\alpha$, $\beta$ and $\gamma$-bands in Sr$_2$RuO$_4$.  
Here we illustrate that the edge mode spectra are considerably affected
by the spin-orbit interaction. 

\begin{figure}[h]
\begin{center}
\includegraphics[width=7cm]{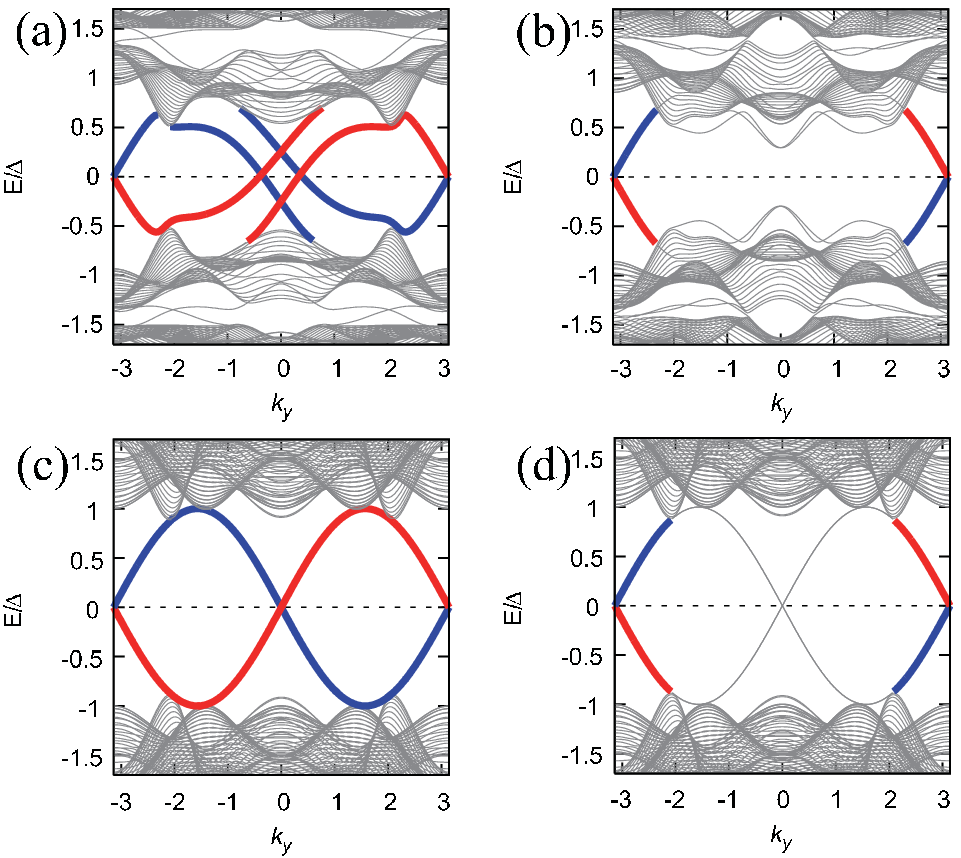}
\caption{ Edge states with [(a) and (b)] and without [(c) and (d)] the
 spin-orbit interaction. 
The ${\bm d}$-vector is chosen as ${\bm d}({\bm k})=\hat{\bm
 z}(\sin k_x+i\sin k_y)$ in (a) and (c), and ${\bm d}({\bm k})=\hat{\bm
 x}\sin k_y-\hat{\bm y}\sin k_x$ in (b) and (d). 
In (c) and (d), the gapless edge modes around $k_y=0$ are two-fold degenerate.
The red (blue) lines represent edge states localized at $x=0$ ($x=L$).
[Around $k_y=0$ in (d), a pair of left and right moving edge modes are
 localized on each edge.] 
We take $\mu_{\rm B}H_z=0$,  $\lambda=0.3$ in (a) and (b), and
 $\lambda=0$ in (c) and (d), respectively.  
Other model parameters are the same as those of Figs.2 (a) and (b).} 
\label{fig:S5}
\end{center}
\end{figure}

In Figs. \ref{fig:S5} (a)-(d), we compare the edge state spectra with
and  without the spin-orbit interaction. 
To sharpen the effect of the spin-orbit interaction, we set the
magnetic field $H_z$ as zero in the numerical calculations in
Fig. \ref{fig:S5}, but other model parameters are the same as those in
Fig. 2.
It is found that the edge state spectra are fairly changed by the
spin-orbit interaction.
In particular, Fig. \ref{fig:S5} clearly indicates that the degeneracy of the
edge modes at $k_y=0$ reported in Refs.[43-47,49]
is resolved by the spin-orbit interaction.

\end{document}